\documentclass[aip,reprint]{revtex4-1}
\usepackage{graphicx}
\usepackage{bm}

\begin{document}

\title{Gravitational interaction of celestial bodies and black holes with particle dark matter} 

\author{S.L.Parnovsky}
\email[]{par@observ.univ.kiev.ua}
\affiliation{Astronomical Observatory, Taras Shevchenko National University of Kyiv, 4 
Observatorna Street, 04053, Kyiv, Ukraine}

\date{\today}

\begin{abstract}
The particles of a dark matter due to gravitational interaction deviate from straight 
trajectories in the vicinity of a massive body. This causes their density to become 
inhomogeneous. The developed density contrast causes a gravitation attraction force acting 
upon the body proportional to its mass squared. Since this effect depends on the body's velocity, 
it causes a specific reference frame to stand out. The 
force is similar to an anisotropic drag, which can be negative in some directions. 
The mean drag force, averaged over all direction, is nonpositive. We can expect some observational 
manifestation of the considered effect for supermassive black holes in galaxies.
\end{abstract}


\maketitle 

\section{Introduction}

The last couple of decades essentially changed our concept about the Universe and its 
contents. The ordinary baryonic matter amounts to 4.6\% of its mean density, 72.6\% corresponds 
to the dark energy and 22.8\% to the dark matter according to the results of data processing of 
{\it Wilkinson Microwave Anisotropy Probe} observations of the cosmic microwave 
background (CMB) radiation anisotropy [\onlinecite{r1}]. 
Astronomers predicted the existence of dark or unseen matter starting from the 1933 
Zwicky paper [\onlinecite{r2}]. Now we have a lot of evidence from astronomy, astrophysics, and 
cosmology in support of this fact. Let us indicate some of them. A large difference between the virial 
masses of galaxy clusters and of the total mass of all galaxies belonging to these clusters 
indicates the presence of dark matter in clusters. Rotation curves of galaxies show that the 
major part of their masses belongs to a dark halo rather than to a stellar disk. We can 
also mention merging galaxy clusters, e.g. the Bullet Cluster 1E0657-558, strong 
and weak gravitational lensing, baryonic acoustic oscillations, simulations of galaxy and 
large-scale structure formation, large-scale collective motion of galaxies, and type Ia supernovae 
explosions. The most important, however, is the CMB anisotropy data, which indicates the 
nonbaryonic nature of the majority of dark matter. 

There are lots of dark matter candidates such as sterile neutrinos, axions, neutralinos, 
gravitinos, and weakly interacting massive particles, which are explained in the 
review [\onlinecite{r3}]. The dark matter must interact gravitationally and cannot take part 
in electromagnetic or strong interactions. It is possible that the dark matter can also 
experience weak interaction. The experiments for the dark matter direct search such as Xenon 
10/100/1T, DAMA/LIBRA, CDMS, CoGeNT, CRESST and many others are trying to detect the dark matter 
assuming that it interacts weakly with ordinary matter. These experiments are
explained in the article [\onlinecite{r4}]. They constrain the cross section of weak interaction
of dark matter to very small values. Nevertheless, there is a possibility that the dark matter 
cannot interact with the ordinary matter via the weak force at all. Sometimes the dark matter, 
which cannot interact weakly is referred to as a mirror matter.

We consider only the gravitational interaction between the particles of the dark matter with 
bodies made from the ordinary matter or black holes. We exclude the weak interaction since we 
are only interested in the effect due to gravity. Naturally, there is 
lot of known effects of the gravitational attraction between such objects and massive aggregates 
of dark matter. We do not consider them. Instead,
we concentrate on the gravitational interaction with the dark matter particles flying through 
and near an object. The considered effect arises due to the trajectory bending of dark matter 
particles in the gravitational field of the object and the gravitational 
attraction of the object to these particles.

We estimate the value of the force and the acceleration caused by this effect and find the 
class of the objects for which this effect  is essential. We consider the cases of isotropic and 
anisotropic velocity distributions of the dark matter particles and show that in both 
cases there is  a specific reference frame, which stands out due to this effect.

\section{Cause of the effect}

Let us start from a preliminary problem. A spherical body (e.g. the Sun 
or any other star) has the mass $M$ and the radius $R$. It rests at the origin of the cylindrical 
coordinate system ($\rho, \phi, z$). A flow of the dark matter passes through and around the body. 
Far from the body at $z \ll -R$ the flow is homogeneous and moves along the $z$ axis with 
the initial velocity $v$. Our goal is to calculate the force acting on the body due to the 
gravitational attraction to dark matter particles.

We consider the case of cold dark matter with the velocity much less than the speed of 
light $v\ll c$ and use the Newtonian mechanics and Newton's law of universal gravitation. 
The masses of dark matter particles are negligibly small as compared to $M$. 

Particles' trajectories deviate from straight lines in the gravitational field of the central 
body. The majority misses the body and moves along hyperbolae. The trajectory of a particle 
deviates by the angle $\psi$
\begin{equation}\label{eqn:1}
\psi=2\arctan \frac{MG}{v^2\rho_0}.
\end{equation}
Here $\rho_0>R$ is an impact parameter and $G$ is the gravitational constant. We assume this 
angle to be small $\psi \ll 1$, so we can assume $\psi\approx 2MG/(v^2\rho_0)$. In this case the 
distance from the center of the body to a particle's periapsis or a point 
of closest approach is almost equal to the impact parameter $\rho_0$.  
As it will be shown further, 
the particles with $\rho_0>R$ provide a lion share of the effect. The particles flying 
through the body with $\rho_0<R$ make a small contribution to it, so we can restrict ourselves 
to a rough estimation. The deviation angle $\psi$ for such particles depends on the density 
distribution of baryonic matter inside the body. We know that at $\rho_0=0$ we have $\psi=0$ 
from the axial symmetry and at $\rho_0=R$ 
this angle must match (\ref{eqn:1}). So we consider an approximation
\begin{equation}\label{eqn:2}                                
\psi=\frac{2MG}{v^2R}\left(\frac{\rho_0}{R}\right)^\gamma,\quad \rho_0<R
\end{equation}
with some positive exponent $\gamma>0$.

\begin{figure}[tb]
\includegraphics[width=\columnwidth]{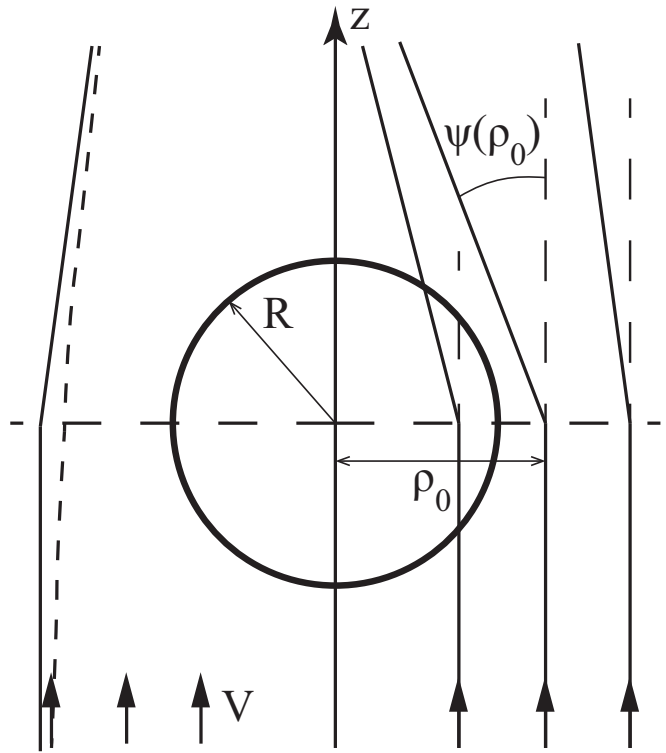}
\caption{\label{fig:1}Dark matter particles in the gravitational field of the spherical body}
\end{figure}
 
In order to demonstrate the physical nature of the effect we consider at first a simple toy model 
(see Fig. \ref{fig:1}). We substitute the real hyperbolic trajectories of dark matter particles 
(the dashed curve in Fig. \ref{fig:1}) by their 
asymptotes. So in the framework of this toy model particles move along straight lines $\rho=\rho_0$ 
parallel to the $z$ axis at $z<0$, at $z=0$ they turn by the angles $\psi(\rho_0)$ and 
continue moving along the straight lines $\rho=|\rho_0-vt\, \sin \psi|, z=vt \, \cos \psi$, where 
$t=0$ corresponds to the moment of crossing the plane $z=0$. These lines intersect at the axis 
$\rho=0$ and then diverge. In addition we consider the motion with constant velocity $v$.
 
The particle density in the half-space $z<0$ is constant. The mass $dm$ passing through the 
area $dS$ of the plane $z$ = const $<0$ during the time interval $dt$ is
\begin{equation}\label{eqn:3}
dm= A dS dt.
\end{equation}
Here $A$ is the flux of the dark matter. 
The penetrating particles occupy the volume $dV= dS\, vdt$, thus the dark matter density 
$\rho_{DM}$  can be expressed as $\rho_{DM}=dm/dV=A/v$. 
In the half-space $z>0$ the distribution of particle density becomes inhomogeneous due to 
the trajectories bending. This leads to the considered effect, namely the difference between 
the gravitational attractions between the body and the particles in both half-spaces. 
To calculate the resultant force let us consider the particles passing through a 
ring-shaped area with $\rho_0<\rho<\rho_0+d\rho$ during the time interval from $t=0$ to $t=dt$. 
At time $t>0$ these particles form a ring with the center at the point $\rho=0, z=vt\cos \psi$. 
The distance between the particles and the center of the body is 
$r=\left( \rho_0^{\phantom{0}2}-2\rho_0vt\, \sin \psi +v^2t^2\right) ^{1/2}$. The mass of the particles 
according to (\ref{eqn:3}) is $dm=\pi A \rho_0 d\rho_0 dt$. The  force due to the gravitational 
attraction between the body and the particles is aligned with the $z$ axis and equals to
\begin{eqnarray}\label{eqn:4}
\begin{array}{l}
\displaystyle F_+=\int \frac{GMzdm}{r^3}=2\pi \rho_{DM}GMv^2\int\limits_0^\infty \cos \psi 
\rho_0 d\rho_0 \\
\displaystyle \phantom{F_+}\times \int\limits_0^\infty
\frac{tdt}{\left( \rho_0^{\phantom{0}2}-2\rho_0vt\, \sin \psi +v^2t^2\right) ^{3/2}}\\
\displaystyle \phantom{F_+}=2\pi \rho_{DM}GM \int\limits_0^\infty \frac{1+\sin \psi}{\cos \psi} d\rho_0
\end{array}
\end{eqnarray}
The force of the gravitational attraction between the body and dark matter in the half-space 
$z<0$ can be obtained in a simple way. It is equal to (\ref{eqn:4}) in the case when the integral is taken 
between the time limits $-\infty$ and 0 at $\psi=0$. Thus its $z$ component $F_-$ is equal 
to $F_-=-F_+(\psi=0)$. The resultant force $F=F_++F_-$. Using (\ref{eqn:4}) we get for the case of 
small deviation angles $\psi$
\begin{eqnarray}\label{eqn:5}
\begin{array}{l}
\displaystyle F=2\pi \rho_{DM}GM \int\limits_0^\infty \psi d\rho_0\\
\displaystyle \phantom{F}=\frac{4\pi \rho_{DM}G^2M^2}{v^2}
\left[ \int\limits_0^R\left(\frac{\rho_0}{R}\right)^\gamma \frac{d\rho_0}{R}
+\int\limits_R^{R_{max}}\frac{d\rho_0}{\rho_0} \right]\\
\displaystyle \phantom{F}=\frac{4\pi \rho_{DM}G^2M^2}{v^2}\left[\frac{1}{1+\gamma}+\ln \frac{R_{max}}{R}\right]\\
\displaystyle \phantom{F}\approx \frac{4\pi \rho_{DM}G^2M^2}{v^2}\ln \frac{R_{max}}{R}.
\end{array}
\end{eqnarray}
This integral diverges at large  $\rho_0$ values, thus we introduced a finite upper limit $R_{max}$. 

The same result can be also obtained straightforwardly without any simplification of the 
calculations. The dark matter particles missing the body move along hyperbolae with varying 
velocities according to the laws of orbital mechanics. We can introduce the useful parameterization 
$t=GM v^{-3}(e\, \cosh  \xi -\xi)$, $r=GMv^{-2}(e\, \cosh  \xi -1)$, $z=r\, 
\sin(\psi /2+\arccos \frac {e-\cosh \xi} {e\, \cosh  \xi -1})$ with deviation angle 
$\psi$ from (\ref{eqn:1}) and the eccentricity of the hyperbola $e=1/\sin (\psi /2)$. After 
integration with respect to $\xi$ between the limits $-\infty$ and $\infty$ we obtain 
for the case $\psi \ll 1$ exactly the expression (\ref{eqn:5}).

Let us note some similarity of this force and the ionization losses of highly energetic charged 
particles or the dynamical friction due to the gravitational interaction of a massive, 
rapidly moving star with a cluster of stars (see [\onlinecite{Lo}]). In all cases $F \propto v^{-2} 
\ln(R_{max}/R_{min})$ and the formula for the dynamical friction coincides with (\ref{eqn:5}). 
Nevertheless, the considered problem differs from the dynamical friction. In the case of the 
dynamical friction we consider a fast body gravitationally interacting with slow ones; in our 
case we consider a slow or immobile body gravitationally interacting with fast particles. 
If these particles move with the same initial velocity, we can use the frame of reference moving 
with this initial velocity and obtain the situation similar to the dynamical friction case. 
Naturally, the force must be the same and this condition is held. But in the general case in 
which particles move in different directions with different velocities we cannot reduce the considered 
problem to the dynamical friction. Later on we will consider this more realistic general case and 
find that the energy of the body can not only decrease, but also increase due to the interaction 
with particles.

\section{Estimation of the value of the force}

The force 
$F$ is proportional to the square of the mass of the body, so it provides an acceleration 
proportional to the mass
\begin{equation}\label{eqn:6}
a= \frac{4\pi \rho_{DM}G^2M}{v^2}\ln \frac{R_{max}}{R}.
\end{equation}

This acceleration is very small because it is proportional to the product of two small 
parameters, namely the gravitational constant squared and the dark matter density. It is 
negligible for ordinary bodies, therefore this effect cannot be found in direct experiment. 
If we draw attention to compact astronomical objects, we shall find a lot of bodies with 
big masses. Moreover, the astronomical time scale provides a lot of time for small but permanent 
acceleration to manifest.

Let us estimate by an order of the magnitude the acceleration of the Sun according to 
(\ref{eqn:6}). Its mass is $M_{Sun}\approx 2\times 10^{30}$ 
kg and the radius $R_{Sun}\approx 7\times 10^{8}$ m. We can estimate $R_{max}$ as the 
half-distance to the nearest stars  $R_{max}\approx  1$ ly $\approx 10^{16}$ m thus 
$\ln (R_{max}/R) \approx  16$. According to the modern paradigm, the dark matter forms a dark 
halo around the Milky Way galaxy and provides the majority of its mass $\approx 10^{12} M_{Sun}$. 
Taking the diameter of the dark halo as 500,000 ly $\approx 5\times 10^{21}$ m we 
can estimate $\rho_{DM}\approx 2\times 10^{-22}$ kg m$^{-3}$. This is an underestimated value 
because of the inhomogeneity of the dark matter distribution. Using the Navarro-Frenk-White 
fitting formula [\onlinecite{r5}] for it 
we can improve this rough estimation. Also we can use the estimations based on astronomical 
observations. A number of $\rho_{DM}$ estimations are cited in the review 
[\onlinecite{r3}]. They vary in the interval 
$0.2 \div 0.8$ GeV cm$^{-3}$. This range was confirmed later by Weber and de Boer 
[\onlinecite{r6}]. Choosing the value 
$\rho_{DM}\approx 0.5$ GeV cm$^{-3}\approx 10^{-21}$ kg m$^{-3}$ and assuming 
$v\approx 200$ km s$^{-1}$ we obtain the estimation $a\approx 2\times 10^{-20}$ m s$^{-2}$. 

This acceleration is less than the constant $a_0$ in the MOND theory by 10 orders of magnitude. 
During the period of the rotation of the Sun around the galactic center 
$\approx 250$ Myr $\approx 8\times 10^{15}$ s this acceleration changes its velocity 
only by 0.16 mm s$^{-1}$. During the age of the Universe 14 Gyr it changes the velocity 
of the Sun by 0.9 cm s$^{-1}$. 

Thus we see that there is a theoretical possibility to transfer the energy, the momentum 
and the angular momentum of the mirror dark matter to the ordinary 
matter. Nevertheless this effect is very small and virtually cannot affect the motion of  
usual celestial bodies. 

Could this effect become significant for other bodies or under different circumstances? We used 
the minimal possible assumption of the value of the typical dark matter particles velocity. 
It corresponds to the Sun's orbital velocity around the center of the Galaxy. Thus we 
cannot decrease the denominator in (\ref{eqn:6}). We have two possibilities to increase the numerator. 
The dark matter density was essentially greater in the early Universe, but there were no 
compact bodies. Thus there remains only one thing to consider -- objects with the mass much greater 
than $M_{Sun}$. Stars have masses up to $\sim 100$ $M_{Sun}$, but the most massive ones have 
lifetimes about a few million years. So the effect is small also for the most massive stars. 
If we consider the acceleration of galaxies we have to use for $\rho_{DM}$ the cosmological 
value $\approx 0.2 \rho_{c}\approx 2\times 10^{-27}$ kg m$^{-3}$, where $\rho_{c}$ is the 
critical cosmological density of the Friedmann universe. Thus, the acceleration is small for 
galaxies as well. 

The most promising objects are supermassive black holes, e.g. the Sagittarius A* black 
hole with the mass $4\times 10^6\,M_{Sun}$  in the center of our Galaxy [\onlinecite{r7}]. Even without 
taking into account the greater dark matter density in the galactic centre we estimate 
$a\approx 8\times 10^{-14}$ m s$^{-2}$. During the period of the rotation of the Sun 
around the galactic center its velocity would change by 640 m s$^{-1}$. Black holes in galaxies, 
e.g. active galactic nuclei (AGN) can be much more massive. The largest supermassive black 
hole in the Milky Way's neighborhood appears to be that of M87 galaxy, weighing about 
$6\times 10^9$ $M_{Sun}$ [\onlinecite{r8}] and the most massive known black hole in the NGC 4889 galaxy 
has the mass $2\times 10^{10}\,M_{Sun}$ [\onlinecite{r9}]. During the period of the rotation of 
the Sun around the galactic center the velocity of the latter changes by 3200 km s$^{-1}$. 
So the considered effect can in principle have some astronomical manifestations, but only for 
supermassive black holes. 

There is another effect, specific for black holes, which arises due to the capture of 
dark matter particles by black holes. We will demonstrate that it is much weaker
than the above mentioned one below.

\section{Isotropic case}

Let us return to the formal problem. There is no reason to assume that the dark matter comes only 
from one direction. We should have used rather a different problem formulation even expecting one 
preferential direction of dark matter motion. In the one-dimensional version of the problem the 
dark matter approaches the body along the direction of the $z$ axis. We suppose that there 
is a frame of reference in which the velocities and the densities of the particles approaching 
from two opposite directions are equal. We denote the particles velocity by $u$. The body is 
moving along the $z$ axis with the velocity $v_0$, $|v_0|<u$. The gravitational force due to the
interaction with dark matter moving along the $z$ axis is directed towards it and 
characterized by (\ref{eqn:5}) with $v=u-v_0$. The gravitational force due to the interaction with dark 
matter moving along the $z$ axis in the negative direction is directed opposite to the first force and 
characterized by (\ref{eqn:5}) with $v=u+v_0$. The resultant force in the direction of the axis is
\begin{eqnarray}\label{eqn:7}
\begin{array}{l}
F=\frac{1}{2}4\pi \rho_{DM}G^2M^2\ln \frac{R_{max}}{R}\left( \frac{1}{(u-v_0)^2}-\frac{1}{(u+v_0)^2}
\right)\\
\displaystyle \phantom{F}=8\pi \rho_{DM}G^2M^2\ln \frac{R_{max}}{R}\frac{uv_0}{(u^2-{v_0}^2)^2}.
\end{array}
\end{eqnarray}
The factor $1/2$ arises because the dark matter density duplicates for the same $A$ due to the 
two possible directions of approaching.
For a small velocity of the body this force is proportional to $v_0$ similar to the drag, 
but with negative viscosity. The value of $v_0$ 
increases with time. At small velocities $|v_0|\ll u$ it increases exponentially. 

Does this effect 
survive if we consider more realistic models in which dark matter particles approach 
from all directions? Let us begin with the simplest one. Let us assume that there is a reference 
frame in which the dark matter flow is isotropic. The particles in this frame have the velocity 
$u$ and their directions are distributed uniformly. Introducing a spherical coordinate 
system with angular coordinates $\theta$ and $\phi$, we can consider the mass of the dark 
matter $dm$ passing through the area $dS$ of some surface during the time interval $dt$ 
from the spatial angle $d\Omega$ in the form similar to the definition of luminance in optics,
\begin{equation}\label{eqn:8}
dm=B(\theta,\phi)\, \cos \beta\, dS\, dt\, d\Omega.
\end{equation}
Here $\beta$ is the angle between the particles' direction of arrival and the direction of 
the surface normal. In the isotropic case $B$=const. It can be easily expressed through the 
dark matter density. The density of the particles with velocity directions in the spatial 
angle $d\Omega$ is $d\rho_{DM}=Bu^{-1} d\Omega$. After integration with respect to $d\Omega$ 
we obtain $\rho_{DM}=4\pi Bu^{-1}$.

\begin{figure}[tb]
\includegraphics[width=\columnwidth]{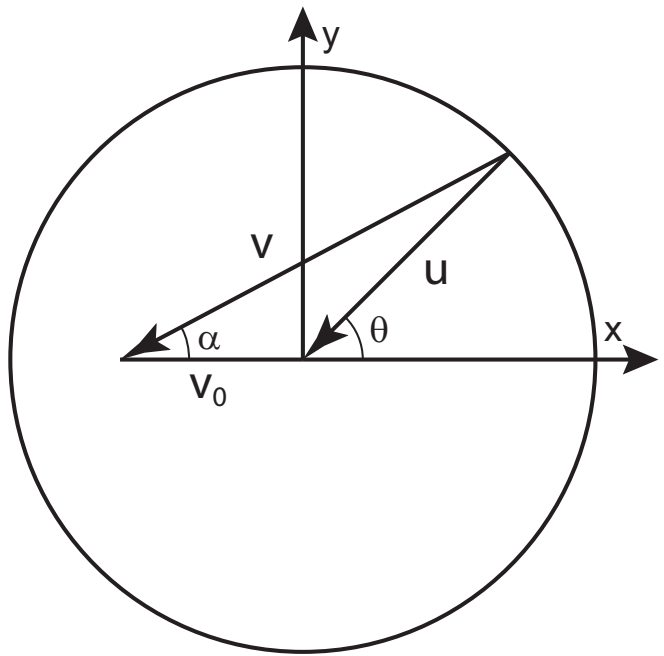}
\caption{\label{fig:2}Relations between velocities and angles in the rest and in the moving frames}
\end{figure}

Particles moving with the velocity $\bm{u}$ from the spatial angle $d\Omega$ provide the gravitational 
force affecting the body in this frame, which we will call the ``rest'' frame,
\begin{eqnarray}\label{eqn:9}
\begin{array}{l}
\displaystyle d\bm{F}=\frac{kd\rho_{DM}}{u^2}\bm{\hat u},\; k=4\pi G^2M^2\ln \frac{R_{max}}{R},\\ 
\displaystyle \bm{\hat u}=\frac{\bm{u}}{u},\; d\rho_{DM}=\frac{B d\Omega}{u}.
\end{array}
\end{eqnarray}
Naturally, after the integration with respect to all directions of the particles arrival we obtain 
$F=0$. 

Let us consider the body moving with the velocity $\bm{v_0}$ with respect to the 
rest frame with the assumption $v_0<u$. We introduce the ratio $\lambda=v_0/u<1$. 
We will use the ``moving'' reference frame in which the body is motionless. 
The velocities of the particles in this system $\bm{v}$ and their components are linked with 
the ones in the rest system $\bm{u}$ by simple relations [\onlinecite{LLM}]
\begin{eqnarray}\label{eqn:10}
\begin{array}{l}
\displaystyle \bm{v}=\bm{u}+\bm{v_0}, v^2=u^2+v_0^{\phantom 0 2}+2uv_0\cos \theta ,\\ 
\displaystyle -v_x=v\,\cos \alpha=v_0+u\, \cos \theta ,\\
-v_y=v\,\sin \alpha=u\, \sin \theta .
\end{array}
\end{eqnarray}
The orientation of  Cartesian coordinates $x,y$ is shown in Fig. \ref{fig:2}. In the moving frame we 
introduced the spherical coordinates $\alpha$ and $\phi$. The angle $\phi$ is the same as in 
the rest frame and the relation between the angles $\theta$ and $\alpha$ follows 
from (\ref{eqn:10}),
\begin{eqnarray}\label{eqn:11}
\begin{array}{l}
\displaystyle \cos \alpha=\frac{\lambda +\cos \theta}{(1+\lambda ^2+2\lambda \cos \theta )^{1/2}},\\
\cos \theta=\cos \alpha \left(1-\lambda ^2\sin ^2\alpha \right)^{1/2}-\lambda \sin ^2\alpha.
\end{array}
\end{eqnarray}
We need the relation between the spatial angle in which the particles come from in the rest 
frame $d\Omega$ and the one the same particles come from in the moving frame, which we 
denote $d\Omega'$ (hereafter all angles with primes correspond to the moving frame). 
Since $d\Omega'=\sin \alpha d\alpha d\phi$ and $d\Omega=\sin \theta d\theta d\phi$, so $d\Omega
=f(\alpha) d\Omega'$ with
\begin{equation}\label{eqn:12}
f(\alpha)=2\lambda \cos \alpha+\frac{1+\lambda^2\cos 2\alpha}{(1-\lambda ^2\sin ^2\alpha)^{1/2}}.
\end{equation}
The particles approaching to the body in the moving frame with the velocity $\bm{v}$ from the 
spatial angle $d\Omega'$ provide the gravitational force affecting the body
\begin{equation}\label{eqn:13}
d\bm{F}=\frac{kd\rho_{DM}}{v^3}\bm{v}, d\rho_{DM}=\frac{B'(\alpha ,\phi) d\Omega'}{v}.
\end{equation}
The value $d\rho_{DM}$ must be the same in any frame of reference. Comparing (\ref{eqn:10}) 
and (\ref{eqn:13}) we obtain
\begin{equation}\label{eqn:14}
B'(\alpha ,\phi)=\frac{B(\theta ,\phi)v}{uf(\alpha)}.
\end{equation}

Let us calculate the force acting on the body in the moving frame in the case when the 
dark matter distribution in the rest frame is isotropic $B(\theta ,\phi)$=const. From the 
axial symmetry it is obvious that this force must be directed along the $x$ axis. So we can 
restrict ourselves to calculating only $F_x$,
\begin{eqnarray}\label{eqn:15}
\begin{array}{l}
\displaystyle F_x= \frac{ku\rho_{DM}}{2}\int\limits_0^\pi \frac{\cos \alpha\, \sin \theta 
d\theta}{v(\theta)^2}\\
\displaystyle \phantom{F_x}=\frac{k\rho_{DM}}{2u}\int\limits_0^\pi \frac{(\lambda 
+\cos \theta) \sin \theta d\theta}{(1+\lambda ^2+2\lambda \cos \theta)^{3/2}}=0.
\end{array}
\end{eqnarray}
Thus in the isotropic case with $v_0<u$ the considered force is absent not only in the rest frame 
but also in any moving frame. 

We can use an analogy with the electrostatics to explain this 
result. Let us consider the model similar to Fig. \ref{fig:2}, but in the coordinate space, not in the velocity 
space. The thin spherical shell with radius $r$ is charged with the distribution of the 
surface charge density $\sigma(\theta,\phi)$ proportional to $B(\theta,\phi)$. If we calculate 
the Coulomb force $F_C$ acting on the test particle located in the point at $x=-\lambda r, y=0$, 
then it will be proportional to the force $F$ due to considered effect. The reason is that 
(\ref{eqn:13}) is similar to the Coulomb's law, but in the velocity space. In the isotropic case 
$\sigma$ = const and we deal with a uniformly charged spherical shell. There is no electric 
field inside such a shell, so there is no force acting on the body in the moving frame.

We can carry on this analogy. Instead of the case when the approaching particles have fixed velocities 
$u$ in the rest frame, let us consider the case where the velocity has some distribution. The mass of the 
dark matter $dm$ passing through the area $dS$ of some surface during the time interval $dt$ 
from the spatial angle $d\Omega$ with the initial velocity $v$ from the interval $u<v<u+du$ is
\begin{equation}\label{eqn:16}
dm=B_u(\theta,\phi,u)\, \cos \beta dSdt d\Omega du.
\end{equation}
In the analogy with the electrostatics $u^3B_u(\theta,\phi,u)$ play the role similar to the 
charge density $\rho_q$. Instead of the thin spherical shell we deal with the charged ball. 
Its charge density $\rho_q(\theta,\phi,r)$ is proportional to $u^3B_u(\theta,\phi,u)$ and $r$ 
is proportional to $u$. Let us consider the isotropic case $B_u(\theta,\phi,u)=B_u(u)$. 
Let the velocities of the approaching particles in the rest frame be distributed in the range 
from $u_{min}$ to $u_{max}$. In the analogy with the electrostatics we deal with spherically 
symmetric charged thick shell with the inner radius $r_{min}\propto u_{min}$ and the outer 
radius $r_{max}\propto u_{max}$. The case $v_0<u_{min}$ corresponds to the configuration in 
which the test particle is located inside the cavity without the  electric field. Thus there is no 
effect in any moving frame with $v_0<u_{min}$ if the motion of the dark matter particles 
is isotropic in the rest frame.

This conclusion completely changes if $v_0>u_{min}$. The force
\begin{equation}\label{eqn:17}
\bm{F}=4\pi k \frac{\int\limits_{u_{min}}^{v_0} B_u(u)u^{-1}du}{{v_0}^3}\bm{v_0}
\end{equation}
acts on the body in the ``moving'' frame. It corresponds to a negative friction. This effect 
takes place only at $v_0>u_{min}$. If $u_{min}=0$ this condition is satisfied automatically for any values.

\section{Anisotropic case}

Let us consider the general case when the angular distribution of arriving 
particles is anisotropic in any reference frame. We start from the arbitrary initial 
reference frame. Once again we utilize the electrostatic analogy and consider the continuous 
distribution of $\rho_q\propto u^3B_u(\theta,\phi,u)\ge 0$ over the velocity space. This charge 
distribution, which we assume to cover a finite area in the velocity space, produces some electric 
field. The distribution of the field intensity over the velocity space must have at least 
one singular point at which its intensity is equal to zero. The existence of such singular 
point is guaranteed by the Brouwer fixed-point theorem for the three-dimensional Euclidean 
space. 

We also can prove an existence of such point from the electrostatic analogue. We have some 
bounded distribution of the positive charge density. The potential of the electric field is 
a continuous function and it decreases outside the charge location, tending to zero at the 
infinity. So, it must be maximal at some point due to the Bolzano–Weierstrass theorem. At this 
point its gradient vanishes and the field intensity is equal to zero. This is a singular point. 
Moreover we can use the Earnshaw's theorem to prove that there is only two possibilities for 
the behavior of the field intensity in the vicinity of any singular point. Either both field 
intensity and the charge density are equal to zero inside some area, or there is some charge 
density in the singular point. We will show below that the later possibility leads to some 
dynamical antidrag force. 

The location of the singular point in the 
velocity space corresponds to a certain velocity. We will refer to the inertial reference frame 
moving with this velocity 
relative to the initial one as a ``special'' frame for the anisotropic case. Its 
location in the velocity space coincides with the singular point. Therefore, the force acting 
on the immobile body in this special frame due to the gravitational interaction with dark 
matter vanishes.

In the isotropic case with nonzero $u_{min}$ the rest frame is surrounded in the velocity 
space by a set of frames in which the resultant force vanishes. All of them could be 
called special frames according to the definition above. Let us verify if in the anisotropic 
case there is such an area of special frames. Using the electrostatic analogy, 
we can reformulate this problem as follows.

There is some static distribution of the electric charge with a cavity inside. Is it possible 
to have no electric field in the cavity in the case of an anisotropic charge distribution? 
To answer this question we use the transformation of inversion $r'=R^2/r$. 
It is known that if some function $U(\bm{r})$ is a solution of 
the Laplace equation $\Delta U=0$, then the function 
$U'(\bm{r'})=R{r'}^{-1}U(R^2\bm{r'}/{r'}^{-2})$ is also its solution [\onlinecite{r10}]. 
We can add an arbitrary constant to the potential. Let us choose the 
value of this constant in such a way that we have $U=0$ inside the cavity and $U=U_{\infty}<0$ 
at the infinity. If we choose an arbitrary point inside the cavity as the origin of the 
coordinate system and perform the inversion, we would turn the charge distribution inside out. Note 
that the point charge $Q=RU_{\infty}<0$ appears at the point $r'=0$. As a result, we get some 
spatially limited charge distribution surrounded by the space without electric field. This means 
that all the terms in the multipole decomposition of the field vanish far from the system of 
charges. It is possible only in the case if the distribution has spherical symmetry or 
consists of spherically-symmetric parts. Thus the initial charge distribution before 
the inversion must also consist from spherically-symmetric parts. The whole charge 
distribution is not necessary spherically symmetric. An example of such distribution is two 
uniformly charged shells one inside another with different centers. Inside the inner shell 
there is no electric field.

Thus an area of the special frames can exist in principle. The distributions of dark matter 
velocities providing an area of the special frames in the velocity space can be thought up. 
But such unusual distributions can hardly be found in the real Universe. Indeed, we can
artificially arrange electrical charge in a special way, but we cannot provide a very special 
distribution of velocities. One can hardly expect to deal with even a distribution with nonzero 
$u_{min}$. For this reason we 
come to the conclusion that the special frame in the anisotropic case is isolated in the 
velocity space.

If the body moves relative to the ``special'' frame, the force due to the gravitational interaction 
with dark matter becomes nonzero. Let us consider the case 
of a small velocity of the body relative to the ``special'' frame. The analogy with the electrostatics 
is useful also for this problem. It is generally known that the electrostatic potential $U$ can be 
expanded into a Taylor series near an arbitrary point used as the origin of the Cartesian 
coordinates $x, y, z$
\begin{equation}\label{eqn:18}
U(x,y,z)=U_0+\sum_{i=1}^3a_ix_i+\sum_{i=1}^3\sum_{i=1}^3 b_{ik}x_i x_k+\cdots.
\end{equation}
Here $x_i$ runs through coordinates $x,y,z$, $U_0=U(0,0,0)$ is the potential at the point, and
vector $\bm{a}$ is the electric field intensity $\bm{E}=-\nabla U$ at the point, i.e at the origin of 
the coordinates, taken with the opposite 
sign. 

If the point is singular, the term with components of $\bm{a}$ vanishes. If we consider the potential distribution near 
the singular point we have to omit this term. The main coordinate-dependent term becomes 
the third one containing a symmetric matrix $b$, which can be reduced to a diagonal form 
$b=\mathrm{diag}(b_1,b_2,b_3)$ by the rotation of the coordinate axes. From the Poisson equation, 
expressed in the Gaussian units, $\Delta U=-4\pi \rho_q$, we get $b_1+b_2+b_3=-2\pi \rho_q \le 0$. 
So, among these values there is at least one negative.
 
Coming back to the considered problem we arrive to the conclusion that in the case when the moving 
frame is located near the special frame in the velocity space and its velocity $\bm{V}$ 
relative to the special frame is small, the components of the anisotropic force acting on 
the body are proportional to the components of $\bm{V}$. 
If we direct the axes of the Cartesian coordinates $x,y,z$ 
along the eigenvectors of the matrix $\tilde b$, which is a velocity-space analog of the matrix 
$b$ appearing in (\ref{eqn:18}), we obtain
\begin{equation}\label{eqn:19}
\bm{F}\propto (\tilde b_1 V_x+\tilde b_2 V_y+\tilde b_3 V_z), \tilde b_1+\tilde b_2+\tilde b_3\ge 0.
\end{equation}
This force is similar to an anisotropic drag. In addition, the viscosity is negative
along at least one eigenvector. The mean friction averaged over all direction is nonpositive.

Let us make the problem more specific. If the body (e.g. the Sun) rotates around the 
galactic center, there are some preferential directions, e.g. the direction towards this center, 
the one perpendicular to the galactic plane, and the one along the direction of the rotation. 
The reference frame associated with the body can be special first of all if the 
density of the particles moving in some direction is equal to the density of particles moving 
in the opposite direction with the same velocity, $B_u(\theta,\phi,u)= B_u(\pi-\theta,\phi+\pi,u)$. 
This means that dark matter's motion is invariant relative to the inversion of time ($T$ invariant). 
This motion can be $T$ invariant if all relaxation processes are finished and if there is no 
rotation of dark matter relative to the galactic rotation. We do not know for sure anything 
concerning the dark matter motion. If dark matter is rotating around the galactic center 
in different way as an ordinary matter does, the motions of stars are affected by small 
accelerations due to the considered effect. As a result, the ordinary matter and dark 
matter subsystems of galaxy can transfer their angular momentum from one to another. But this
effect is small and practically does not affect the galactic dynamics.

If the velocity of the special frame relative to the frame of the CMB radiation isotropy 
is coherent with the rotation of the part of galaxy  
for which this frame is special, then the special frames for different galaxies or 
different parts of galaxy are rotating relative to each other. In this case there is no inertial 
frame, which is special for the whole Universe. If such a universal inertial special 
frame exists, it means that 
the dark matter subsystems of all galaxies do not rotate around their centers and the angular 
momenta of these subsystems are equal to zero.

\section{Capture of dark matter particles by black holes}

The estimation made in Sec, 3 shows that the considered effect is significant only for supermassive 
black holes. As mentioned earlier, there is another effect acting in this case. Previously, we 
considered the dark 
matter without weak interaction to avoid any nongravitational interaction with the body. 
All particles penetrating the body escape it, as it is shown in Fig. \ref{fig:1}. But particles 
penetrating the black hole horizon cannot escape. They transfer their momentum to 
the black hole providing a force acting on it. We need a formula for the black hole  
capture cross section $\sigma_0$ for the cold dark matter to calculate this force. This 
problem was considered many times (see, e.g. [\onlinecite{r11}], problem 15.11) and its 
solution is
\begin{equation}\label{eqn:20}
\sigma_0=\frac{16\pi G^2M^2}{c^2u^2}, u\ll c.
\end{equation}
It is easy to calculate the corresponding force
\begin{equation}\label{eqn:21}
d\bm{F}=\frac{16\pi G^2M^2}{c^2}d\rho_{DM}\bm{\hat u}.
\end{equation}
Comparing this equation with (\ref{eqn:9}) we see that the force (\ref{eqn:21}) is much 
less that (\ref{eqn:9}), so we can neglect it.

From (\ref{eqn:20}) one can see that a minimal impact parameter $\rho_{max}$ for avoiding 
the capture of the dark matter particles is proportional to the Schwarzschild radius $r_g$ of 
the black hole $\rho_{min}=2r_g\frac{c}{u}$. For our estimation of the typical value of dark 
matter particles velocity we have $\rho_{min}\approx 3\times 10^{3}r_g$. The deviation angle 
$\psi\ll 1$ if $\rho\gg \rho_1=r_g\left( \frac{c}{u}\right)^2$. If we assume that all dark 
matter particles with the impact parameter $\rho_{max}<\rho_1$ transfer their momentum to 
the black hole, this provide the force
\begin{equation}\label{eqn:22}
d\bm{F}=\frac{4\pi G^2M^2}{u^2}d\rho_{DM}\bm{\hat u}
\end{equation}
acting on the black hole. Note that it is less that the force (\ref{eqn:5}). The real force 
acting on the black hole due to the gravitational interaction with the particles with the 
impact parameters $\rho_{min}<\rho<\rho_1$ is less that (\ref{eqn:22}).  Thus the lion share 
of the force acting on the supermassive black hole provide particles moving on the distances 
more that $\rho_1$, where the assumption $\psi\ll 1$ holds. Also holds the weak gravitational 
field approximation and we can use the Newtonian mechanics.

The major part of the force is provided by the particles missing the body or the black hole on 
distances from  about 10000 Schwarzschild radii to an $R_{max}$ many times more. In this region 
around the black hole in the real Universe could be an accretion disk or some orbiting compact 
objects. The gravitational interaction with them can interfere with the gravitational attraction 
to the central massive black hole, but the difference in scales of the effects probably cannot 
allow the essential deviation from the formula (\ref{eqn:5}). 
The ordinary matter particles also can contribute to this effect. But they can interact 
additionally with the matter and the electromagnetic field surrounding the body. Ordinary 
matter cannot pass through the accretion disc around the black hole without nongravitational 
interaction, but the mirror dark matter can.  However, the interaction of the ordinary matter 
approaching the body body is much harder to study that the practically pure gravitational 
interaction of the dark matter particles.

\section{Conclusion}

We considered the force acting on a body or a black hole due to the bending of dark matter 
particles' trajectories in the gravitational field of this massive object. This force 
depends on the velocity of the object.
The force is also proportional to the body's mass squared, and the acceleration caused by it is 
proportional to the mass of the body. The estimation of its value shows that it 
is very small even for stars. But there are objects for which we can expect some observational 
manifestation of the considered effect, especially on the cosmological time scale. These are the
supermassive black holes in galaxies, including active
galactic nuclei. But the considered effect can be masked 
by the interaction of the ordinary matter with other baryonic matter or the electromagnetic field 
surrounding the black hole.

Moreover, it can be observed only in the case of essential anisotropy in the distribution
of velocities of dark matter particles in the frame of the black hole. This effect can manifest in
a large velocity of a supermassive black hole relative to other objects in the vicinity. We cannot 
measure the tangential components of the velocity and the effect can show itself as a difference
in redshifts of a black hole and a part of galaxy near it. If such a difference was detected 
for supermassive black holes with masses greater than $10^9$ $M _{Sun}$ only, this would be an 
argument in favor of the existence of the considered effect.  

This force allows us to select a preferable reference frame.
In any point of the space there is a special inertial frame of reference in which 
there is no resultant force acting on the immobile body due to gravitational interaction with 
the dark matter particles flying near and through the body. If the distribution of the 
dark matter velocity is anisotropic in this frame, then in the frames moving steady 
and straight relative to this special frame there are the small forces acting on 
the bodies, which are immobile in these frames. In other words, the force is acting on the 
moving bodies in the special frame. For the small body velocities in the special 
frame, the force is proportional to the velocity components in the way similar to the case 
of an anisotropic drag force (\ref{eqn:19}). There is always a directions of motion with a 
negative viscosity coefficient. The mean friction averaged over all direction is nonpositive.

If the distribution of the dark matter velocity is isotropic in some special frame 
and the minimal velocity of the particles approaching far from the body $u_{min}$ is equal 
to zero, the force acting on the body moving relative to this special frame with the 
velocity $\bm{v}$ is directed towards $\bm{v}$. This case is similar to an isotropic drag
with a negative viscosity. If $u_{min}\ne 0$, then there are many special frames. 
The preferable one is the rest system in which the distribution of the dark matter 
velocities is isotropic. All the frames with the velocities $v<u_{min}$ relative to the rest 
frame are special ones. In other words, the is no force acting on bodies moving 
with velocities $v<u_{min}$ in the rest frame. If the body's velocity exceeds $u_{min}$, 
the force acts towards the direction of motion.  

Nowadays we know some preferable inertial frames, e.g. the frame of CMB isotropy. To a certain 
degree we return to Aristotle's time. At any location in the Universe there are inertial frames 
particular for considered effect. This is the special frame in the case of anisotropy of 
dark matter distribution or the rest frame for the isotropic case. 
These frames are local ones. If there is a universal inertial frame, which is special for 
the whole Universe, it means that the angular momenta of dark matter subsystems of all 
galaxies are equal to zero.





\bibliography{parn}

\providecommand{\noopsort}[1]{}\providecommand{\singleletter}[1]{#1}%
\begin{thebibliography}{13}%
\makeatletter
\providecommand \@ifxundefined [1]{%
 \@ifx{#1\undefined}
}%
\providecommand \@ifnum [1]{%
 \ifnum #1\expandafter \@firstoftwo
 \else \expandafter \@secondoftwo
 \fi
}%
\providecommand \@ifx [1]{%
 \ifx #1\expandafter \@firstoftwo
 \else \expandafter \@secondoftwo
 \fi
}%
\providecommand \natexlab [1]{#1}%
\providecommand \enquote  [1]{``#1''}%
\providecommand \bibnamefont  [1]{#1}%
\providecommand \bibfnamefont [1]{#1}%
\providecommand \citenamefont [1]{#1}%
\providecommand \href@noop [0]{\@secondoftwo}%
\providecommand \href [0]{\begingroup \@sanitize@url \@href}%
\providecommand \@href[1]{\@@startlink{#1}\@@href}%
\providecommand \@@href[1]{\endgroup#1\@@endlink}%
\providecommand \@sanitize@url [0]{\catcode `\\12\catcode `\$12\catcode
  `\&12\catcode `\#12\catcode `\^12\catcode `\_12\catcode `\%12\relax}%
\providecommand \@@startlink[1]{}%
\providecommand \@@endlink[0]{}%
\providecommand \url  [0]{\begingroup\@sanitize@url \@url }%
\providecommand \@url [1]{\endgroup\@href {#1}{\urlprefix }}%
\providecommand \urlprefix  [0]{URL }%
\providecommand \Eprint [0]{\href }%
\providecommand \doibase [0]{http://dx.doi.org/}%
\providecommand \selectlanguage [0]{\@gobble}%
\providecommand \bibinfo  [0]{\@secondoftwo}%
\providecommand \bibfield  [0]{\@secondoftwo}%
\providecommand \translation [1]{[#1]}%
\providecommand \BibitemOpen [0]{}%
\providecommand \bibitemStop [0]{}%
\providecommand \bibitemNoStop [0]{.\EOS\space}%
\providecommand \EOS [0]{\spacefactor3000\relax}%
\providecommand \BibitemShut  [1]{\csname bibitem#1\endcsname}%
\let\auto@bib@innerbib\@empty
\bibitem [{\citenamefont {Komatsu}\ \emph {et~al.}(2009)\citenamefont {Komatsu}
  \emph {et~al.}}]{r1}%
  \BibitemOpen
  \bibfield  {author} {\bibinfo {author} {\bibfnamefont {E.}~\bibnamefont
  {Komatsu}} \emph {et~al.},\ }\href@noop {} {\bibfield  {journal} {\bibinfo
  {journal} {Astrophys. J.}\ }\textbf {\bibinfo {volume} {180}},\ \bibinfo
  {pages} {330} (\bibinfo {year} {2009})}\BibitemShut {NoStop}%
\bibitem [{\citenamefont {Zwicky}(1933)}]{r2}%
  \BibitemOpen
  \bibfield  {author} {\bibinfo {author} {\bibfnamefont {F.}~\bibnamefont
  {Zwicky}},\ }\href@noop {} {\bibfield  {journal} {\bibinfo  {journal}
  {Helvetica Phys. Acta}\ }\textbf {\bibinfo {volume} {6}},\ \bibinfo {pages}
  {110} (\bibinfo {year} {1933})}\BibitemShut {NoStop}%
\bibitem [{\citenamefont {Bertone}, \citenamefont {Hooper},\ and\ \citenamefont
  {Silk}(2005)}]{r3}%
  \BibitemOpen
  \bibfield  {author} {\bibinfo {author} {\bibfnamefont {G.}~\bibnamefont
  {Bertone}}, \bibinfo {author} {\bibfnamefont {D.}~\bibnamefont {Hooper}}, \
  and\ \bibinfo {author} {\bibfnamefont {J.}~\bibnamefont {Silk}},\ }\href@noop
  {} {\bibfield  {journal} {\bibinfo  {journal} {Phys. Rep.}\ }\textbf
  {\bibinfo {volume} {405}},\ \bibinfo {pages} {279} (\bibinfo {year}
  {2005})}\BibitemShut {NoStop}%
\bibitem [{\citenamefont {Szelc}(2010)}]{r4}%
  \BibitemOpen
  \bibfield  {author} {\bibinfo {author} {\bibfnamefont {A.~M.}\ \bibnamefont
  {Szelc}},\ }\href@noop {} {\bibfield  {journal} {\bibinfo  {journal} {Acta
  Phys. Polon. B}\ }\textbf {\bibinfo {volume} {41}},\ \bibinfo {pages} {1417}
  (\bibinfo {year} {2010})}\BibitemShut {NoStop}%
\bibitem [{\citenamefont {Longair}(2011)}]{Lo}%
  \BibitemOpen
  \bibfield  {author} {\bibinfo {author} {\bibfnamefont {M.~S.}\ \bibnamefont
  {Longair}},\ }\href@noop {} {\emph {\bibinfo {title} {High Energy
  Astrophysics}}}\ (\bibinfo  {publisher} {Cambridge University Press,
  Cambridge},\ \bibinfo {year} {2011})\BibitemShut {NoStop}%
\bibitem [{\citenamefont {Navarro}, \citenamefont {Frenk},\ and\ \citenamefont
  {White}(1997)}]{r5}%
  \BibitemOpen
  \bibfield  {author} {\bibinfo {author} {\bibfnamefont {J.~F.}\ \bibnamefont
  {Navarro}}, \bibinfo {author} {\bibfnamefont {C.~S.}\ \bibnamefont {Frenk}},
  \ and\ \bibinfo {author} {\bibfnamefont {S.~D.~M.}\ \bibnamefont {White}},\
  }\href@noop {} {\bibfield  {journal} {\bibinfo  {journal} {Astrophys. J.}\
  }\textbf {\bibinfo {volume} {490}},\ \bibinfo {pages} {493} (\bibinfo {year}
  {1997})}\BibitemShut {NoStop}%
\bibitem [{\citenamefont {Weber}\ and\ \citenamefont {de~Boer}(2010)}]{r6}%
  \BibitemOpen
  \bibfield  {author} {\bibinfo {author} {\bibfnamefont {M.}~\bibnamefont
  {Weber}}\ and\ \bibinfo {author} {\bibfnamefont {W.}~\bibnamefont
  {de~Boer}},\ }\href@noop {} {\bibfield  {journal} {\bibinfo  {journal}
  {Astron. Astrophys.}\ }\textbf {\bibinfo {volume} {509}},\ \bibinfo {pages}
  {id.A25} (\bibinfo {year} {2010})}\BibitemShut {NoStop}%
\bibitem [{\citenamefont {Ghez}\ \emph {et~al.}(2008)\citenamefont {Ghez} \emph
  {et~al.}}]{r7}%
  \BibitemOpen
  \bibfield  {author} {\bibinfo {author} {\bibfnamefont {A.~M.}\ \bibnamefont
  {Ghez}} \emph {et~al.},\ }\href@noop {} {\bibfield  {journal} {\bibinfo
  {journal} {Astrophys. J.}\ }\textbf {\bibinfo {volume} {689}},\ \bibinfo
  {pages} {1044} (\bibinfo {year} {2008})}\BibitemShut {NoStop}%
\bibitem [{\citenamefont {Gebhardt}\ and\ \citenamefont {Thomas}(2009)}]{r8}%
  \BibitemOpen
  \bibfield  {author} {\bibinfo {author} {\bibfnamefont {K.}~\bibnamefont
  {Gebhardt}}\ and\ \bibinfo {author} {\bibfnamefont {J.}~\bibnamefont
  {Thomas}},\ }\href@noop {} {\bibfield  {journal} {\bibinfo  {journal}
  {Astrophys. J.}\ }\textbf {\bibinfo {volume} {700}},\ \bibinfo {pages} {1690}
  (\bibinfo {year} {2009})}\BibitemShut {NoStop}%
\bibitem [{\citenamefont {McConnell}\ \emph {et~al.}(2011)\citenamefont
  {McConnell} \emph {et~al.}}]{r9}%
  \BibitemOpen
  \bibfield  {author} {\bibinfo {author} {\bibfnamefont {N.~J.}\ \bibnamefont
  {McConnell}} \emph {et~al.},\ }\href@noop {} {\bibfield  {journal} {\bibinfo
  {journal} {Nature}\ }\textbf {\bibinfo {volume} {480}},\ \bibinfo {pages}
  {215} (\bibinfo {year} {2011})}\BibitemShut {NoStop}%
\bibitem [{\citenamefont {Landau}\ and\ \citenamefont {Lifshitz}(1976)}]{LLM}%
  \BibitemOpen
  \bibfield  {author} {\bibinfo {author} {\bibfnamefont {L.~D.}\ \bibnamefont
  {Landau}}\ and\ \bibinfo {author} {\bibfnamefont {E.~M.}\ \bibnamefont
  {Lifshitz}},\ }\href@noop {} {\emph {\bibinfo {title} {Mechanics}}}\
  (\bibinfo  {publisher} {Butterworth-Heinemann, Oxford},\ \bibinfo {year}
  {1976})\BibitemShut {NoStop}%
\bibitem [{\citenamefont {Landau}, \citenamefont {Lifshitz},\ and\
  \citenamefont {Pitaevskii}(1984)}]{r10}%
  \BibitemOpen
  \bibfield  {author} {\bibinfo {author} {\bibfnamefont {L.~D.}\ \bibnamefont
  {Landau}}, \bibinfo {author} {\bibfnamefont {E.~M.}\ \bibnamefont
  {Lifshitz}}, \ and\ \bibinfo {author} {\bibfnamefont {L.~P.}\ \bibnamefont
  {Pitaevskii}},\ }\href@noop {} {\emph {\bibinfo {title} {Electrodynamics of
  Continuous Media}}}\ (\bibinfo  {publisher} {Butterworth-Heinemann},\
  \bibinfo {year} {1984})\BibitemShut {NoStop}%
\bibitem [{\citenamefont {Lightman}\ \emph {et~al.}(1975)\citenamefont
  {Lightman}, \citenamefont {Press}, \citenamefont {Price},\ and\ \citenamefont
  {Teukolsky}}]{r11}%
  \BibitemOpen
  \bibfield  {author} {\bibinfo {author} {\bibfnamefont {A.~P.}\ \bibnamefont
  {Lightman}}, \bibinfo {author} {\bibfnamefont {W.~H.}\ \bibnamefont {Press}},
  \bibinfo {author} {\bibfnamefont {R.~H.}\ \bibnamefont {Price}}, \ and\
  \bibinfo {author} {\bibfnamefont {S.~A.}\ \bibnamefont {Teukolsky}},\
  }\href@noop {} {\emph {\bibinfo {title} {Problem Book in Relativity and
  Gravitation}}}\ (\bibinfo  {publisher} {Princeton University, Princeton},\
  \bibinfo {year} {1975})\BibitemShut {NoStop}%
\end{thebibliography}%
\end{document}